# Insight into Ideal Shear Strength of Ni-based Dilute Alloys using First-Principles Calculations and Correlational Analysis

John D. Shimanek[1*], Shun-Li Shang[1,*], Allison M. Beese[1,2], and Zi-Kui Liu[1]

[1] *Department of Materials Science and Engineering, The Pennsylvania State University, University Park, PA, 16802, USA*

[2] *Department of Mechanical Engineering, The Pennsylvania State University, University Park, PA 16802, USA*

* *Corresponding authors: jshimanek@psu.edu, sus26@psu.edu*

**Abstract:**

The present work examines the effect of alloying elements (denoted X) on the ideal shear strength for 26 dilute Ni-based alloys, $Ni_{11}X$, as determined by first-principles calculations of pure alias shear deformations. The variations in ideal shear strength are quantitatively explored with correlational analysis techniques, showing the importance of atomic properties such as size and electronegativity. The shear moduli of the alloys are affirmed to show a strong linear relationship with their ideal shear strengths, while the shear moduli of the individual alloying elements were not indicative of alloy shear strength. Through combination with available ideal shear strength data on Mg alloys, a potential application of the Ni alloy data is demonstrated in the search for a set of atomic features suitable for machine learning applications to mechanical properties. As another illustration, the calculated Ni ideal shear strengths play a key role in a predictive multiscale framework for deformation behavior of single crystal alloys at large strains, as shown by simulated stress-strain curves.

## 1 Introduction

The ideal shear strength of a material is the stress required to simultaneously shear all bonds along a given crystallographic plane. This situation represents the most difficult way to accomplish shear deformation, and the value of a material's ideal shear strength correspondingly sets an upper bound on its strength [1,2]. Importantly, this upper bound can be predicted by first-principles calculations based on density functional theory (DFT) [3] and compared with small-scale experimental results of deformation, where specimen geometry prohibits the normal deformation mechanisms via dislocations and allows stresses to approach the ideal strength [4–6]. Common values between such disparate length scales are rare, especially for mechanical properties, which are influenced by many extrinsic factors. By capitalizing on its intrinsic nature, the ideal shear strength can serve as a linkage across the length scales of material strength calculations, as recently demonstrated for the hardening behavior of pure Ni single crystals [7].

As early as 1926, the ideal shear stress was calculated by Frenkel to be

$$\tau_{is} = \frac{Gb}{2\pi h} \tag{1}$$

where $G$ is the shear modulus of the material, $b$ is the Burgers vector, and $h$ is the spacing between slip planes [1,8]. Much later, such analytical forms were largely borne out by first-principles calculations [3]. This level of stress is difficult to realize in bulk specimens due to the existence of dislocations that accomplish deformation at much lower stresses by allowing bonds in the shear plane to break and reform sequentially. However, as sample size decreases below around 10μm, material strengths begin to approach ideal values due to the spatial constraints on dislocations and their sources [9,10]. As a result, the ideal shear strength can be probed with nanoindentation [4,5] and micropillar compression [6], although the values may remain equivocal due to experimental challenges such as complex stress states [5] or ion-implantation effects [11]. Despite the difficulties of precisely resolving the ideal shear strength of materials experimentally, the recent ability of experiments to approach these values has meant that experimental and computational methods share the ideal shear strength as a common point of comparison.

The ideal shear strength, obtained through first-principles calculations [12] or atomistic simulations [13], could be used to inform alloy design. The most complete examples of multiscale frameworks for the prediction of mechanical behavior include PRISMS [14] and DAMASK [15], neither of which currently utilizes ideal shear strength as a parameter due to its unestablished connection to their included constitutive equations. The prediction of constitutive behavior from first-principles calculations represents a missing capability in the suite of available computational tools for alloy design, which includes such established techniques as the calculation of phase diagram (CALPHAD) and phase field methods to predict phases and microstructures and the crystal plasticity finite element method (CPFEM) to calculate deformation under different loading conditions [16–18].

Motivated by the key role a predictive framework of metal plasticity would fill in the overall process of computational alloy design, recent progress has demonstrated the utility of first-principles ideal shear strength calculations in their application to the strain hardening behavior of pure Ni [7]. In these calculations, an additional minor strain component was imposed on the supercell to mimic long-range strain fields, which can influence the motion of dislocations. The resulting increase in ideal shear strength was translated, through the theoretical framework of the Peierls-Nabarro equation [19], into an increase in the respective flow resistances for pure edge and pure screw dislocations. A linear model was proposed to combine contributions from both edge and screw dislocation segments as a function of strain, to account for the formation of dislocation junctions, which resulted in good agreement between CPFEM results and experiments at large strains and different orientations of pure Ni single crystals [7].

As pure metals themselves are of limited engineering use, an extension of the framework presented in Ref. [7] is necessary to incorporate the effects of alloying elements. However, despite the ideal shear strength of pure elements routinely being the subject of first-principles calculations, the relationship between substitutional solid solution alloying elements and ideal shear strength in face-centered cubic (fcc) metals is rarely explored, except as a means to examine the unique case of complex concentrated alloys [20,21]. The literature contains several studies on the effect of substitutional solutes on the ideal shear strength of specific alloy systems, such as Cr in body-centered cubic Fe [22], alloying elements in Mg [23], and a few key alloying elements in Ti [24]. Studies on alloying effects in Ni are more often undertaken for one of its strengthening intermetallic phases, $Ni_3Al$, than for the fcc matrix phase [25,26]. Advancement toward a predictive computational framework for the strength of materials requires a thorough exploration of the changes in ideal shear strength with respect to alloying elements.

Recently, machine learning methods have emerged as one of the most promising approaches to explore complex structure-property relationships in materials science. Machine learning techniques have been applied to a wide range of materials research activities [27], including design of metallic glasses [28], inversion of microstructure-property relations [29], estimation of alloy machinability [30], and prediction of solute segregation energy spectra in polycrystals [31]. Despite their potential, machine learning methods rely on abundant data for model training. In the case of mechanical properties, experimental data can have a very high dimensionality – e.g., composition, synthesis method, numerous microstructural features, surface quality – but a very low total number of observations. First-principles methods promise ample training data while at the same time limiting the scope of the input data to atomic properties. Machine learning has already been used in concert with data produced from first-principles methods for the case of ternary oxides [32], stacking fault energy in dilute fcc-based alloys [33], and, more generally, its integration with first-principles calculations in the generation of thermodynamic data is an active research topic [34]. However, there is not yet an established list of atomic features important for machine learning applications to the mechanical behavior of materials. One perspective through which to view feature selection for mechanical properties is that of the ideal shear strength. The ideal shear strength can be calculated using first-principles methods to generate sufficient data, yet it describes a real measure of material strength. In the present study, atomic features are incorporated into the investigation of ideal shear strengths of dilute Ni alloys to explore the association between each alloying element property and the resulting ideal shear strength of the alloy.

The present study aims to investigate the ideal shear strengths of 26 dilute Ni-X binary alloys via first-principles techniques and to explore which attributes of the alloying elements are important to the ideal shear strength of the alloy. Additionally, trends in the ideal shear strength of the alloys are explored in the context of feature selection, where multiple measures of association are used to investigate the correlations between the calculated ideal shear strength and several elemental features. To build a more general feature set for mechanical property predictions via correlation analysis techniques, the Ni-X ideal shear strength data must be combined with similar data from other alloy systems, which is demonstrated for the case of dilute Mg-based alloys. Finally, the Ni-X ideal shear strengths serve to extend a recently-proposed method to predict macroscopic single crystal deformation behavior [7] and to investigate the trends in deformation behavior.

## 2 Methods

2.1 Ideal shear strength from pure alias shear deformation

DFT-based first-principles calculations were performed to determine the ideal shear strength of 26 Ni-X binary alloys with X=Al, Si, Sc, Ti, V, Cr, Mn, Fe, Co, Cu, Zn, Y, Zr, Nb, Mo, Tc, Ru, Rh, Pd, Hf, Ta, W, Re, Os, Ir, and Pt. These elements were chosen as some of the most common alloying elements of Ni-based alloys [35], and, additionally, some elements were added from the first three rows of transition metal elements to better capture trends across the periodic table. Each of the alloying elements replaces one Ni atom within the shear plane of the 12-atom supercell, as shown in Figure 1. A rotated version of the conventional fcc lattice, i.e., a 12-atom orthorhombic cell ($\alpha = \beta = \gamma = 90°$), was used to facilitate the resolution of stresses and strains onto the $\{111\}\langle 11\bar{2}\rangle$ partial slip system orientation. The base plane of the orthorhombic cell represents the $(111)$ plane and its axes the $[11\bar{2}]$ and $[\bar{1}10]$ directions. It contains 3 atomic layers in the $[111]$ direction – more than this minimum number of layers reduces the calculated ideal shear strength, bringing it further from experimental values, as discussed in Ref. [7]. The initial lattice parameters in the $[11\bar{2}]$, $[\bar{1}10]$, $[111]$ directions were 4.309, 4.973, and 6.099 Å [7,36,37], respectively, where these initial values do not affect the resulting ideal shear strengths obtained by constrained relaxations, as detailed below. With this configuration, alias shear deformation in the $\{111\}\langle 11\bar{2}\rangle$ partial slip system can be applied to the angle between two axes through the deformation tensor $\boldsymbol{F}$:

$$\boldsymbol{F} = \begin{bmatrix} 1 & 0 & 0 \\ 0 & 1 & 0 \\ \varepsilon & 0 & 1 \end{bmatrix} \tag{2}$$

where $\varepsilon$ is the magnitude of shear strain. This type of deformation along the partial dislocation system represents the softest ideal deformation mode of the crystal when $\varepsilon$ is positive. The deformation was imposed throughout relaxations of atomic positions and non-prescribed cell parameters by an external optimization program, Gadget [38]. All relaxed stress components were less than 0.15 GPa except for the imposed shear, and forces acting on atoms were below 0.03 eV/Å. The engineering shear strain was calculated as the shear displacement divided by the height of the supercell.

First-principles studies of the ideal shear strength of metals are regularly performed using a total energy approach based on DFT. For instance, to find the generalized stacking fault energy of Mg-based alloys with Li and Al, Han et al. [39] calculated the energy of the system as two slabs of the material slid against each other. The maximum derivative of the resulting energy curve in the softest direction was taken as the ideal shear strength, as described earlier by Ogata et al. [12] for pure Cu and Al. However, the slab deformation approach uses a simple shear deformation, where relaxation is allowed only normal to the shear plane. In contrast, the deformation mode used in this work is that of pure alias shear [40]. Here, "pure" indicates a relaxation of both atomic positions and cell shape, except for the fixed shear angle (as opposed to the unrelaxed case of simple shear), and "alias" refers to the deformation regime where only one atomic layer is displaced (in contrast to affine shear, wherein all atomic rows are displaced proportionally to their distance from the shear plane) [12,40]. Due to the relaxation during pure alias shear, the deformation of one layer of the material naturally propagates through the rest of the atomic layers.

The value of the applied shear strain was incrementally increased, and the associated shear stress was recorded for each converged calculation. The shear stress increased as a function of strain up to a drop in stress that was often sharp but sometimes smooth, as discussed in Section 3.1. Regardless of the abruptness of the stress drop, the maximum stress was taken to be the ideal shear strength. After overcoming the Peierls barrier, i.e., beyond the zero-stress point corresponding to an unstable stacking fault, the direction of stress inverts so that a negative stress still pushes the shear plane away from the point of maximum stacking fault energy. The procedure of geometry optimization for each of several imposed shear strains was repeated for 26 supercells, each containing a single atom of a different alloying element. Once the approximate shear strain of the maximum stress and stress drop was realized, several more calculations were added around this critical area of the datasets to ensure a precise ideal shear strength, converged to within 0.08 GPa.

### 2.2 First-principles calculations

First-principles calculations were performed using the Vienna *Ab initio* Simulation Package (VASP) [41] and the projector augmented wave method [42]. The chosen exchange-correlation functional was the generalized gradient approximation as parameterized by Perdew et al. [43]. An energy cutoff of 350 eV was used along with a Γ-centered *k*-point mesh of 9×8×7 points. Core configurations were chosen based on VASP's recommendations, and they are the same as previously used in similar calculations [44,45]. The energy convergence criterion for electronic self-consistency was $5\times10^{-6}$ eV per supercell. A Methfessel-Paxton reciprocal space energy integration scheme was used with a smearing width of 0.2 eV [46].

Following the procedure introduced in Ref. [47], the elastic constants were calculated for pure Ni and all 26 $Ni_{11}X$ alloys. Fixed non-zero strains of ±0.007 and ±0.013 were applied to 6 deformation modes (3 normal and 3 shear), and the resulting forces were used to construct stiffness matrices. The stiffness matrices from both strains were averaged by component to get a final stiffness matrix for each alloy studied, which was used to find an effective polycrystalline shear modulus using the Hill method of averaging the Voigt and Reuss limits [48].

## 2.3 Correlational analysis and feature selection

To understand the underlying connections between the physical properties of the alloying elements and the resulting ideal shear strength of the alloy, the influence of each alloying element was examined through correlational analysis and feature selection algorithms. Features were chosen for being material-specific physical properties, available in the literature for all alloying elements considered in the present work without additional calculations. Since the structure of each alloy is the same fcc Ni lattice, there were no structural features included. To account for the effect of the host element, and in accordance with previous literature on substitutional solid solutions [33,49], the difference of each feature value from the host element value was used instead of absolute feature values. The features are listed in Table 1 and grouped in several broad categories of elemental attributes [33]:

- Atomic: covalent and Van der Waals radii, mass, and volume
- Periodic: number, period and group in the periodic table, and Mendeleev number
- Elastic: Young's, shear, and bulk moduli
- Thermodynamic: melting and boiling points, heats of fusion and sublimation, electronic and heat conductivity, heat capacity, and standard entropy at 298K
- Lattice: cohesive energy, Debye temperature, vacancy formation and activation energies; and
- Electronic: electron density and affinity, electronegativity, ionization potentials, pseudopotential radii of the s and p orbitals, number of spectral lines, maximum range of electrons in solids, and number of filled/unfilled electron states in each shell.

Several techniques from correlational analysis can give insightful perspectives on data with high dimensionality, even on the relatively small datasets common to materials science problems. These feature selection techniques can be broadly categorized as filter-type, wrapper-type, or embedded methods [50]. Filter-type methods, such as the coefficient of determination (i.e., $R^2$) often used in linear regression, require no machine learning model at all and are governed by deterministic algorithms. Wrapper-type methods use a machine learning model to evaluate the performance of many subsets of features, eventually deciding on an optimal list. Lastly, embedded methods are used within the training of a machine learning algorithm to apply a penalty to each additional feature included, limiting their total number in the final model; embedded methods were not used in the present investigation since their selection of predictive features depends heavily on the choice of model [50].

Filter methods include the maximal information coefficient (MIC), F-tests, and the Relief algorithm along with the standard coefficient of determination. F-tests extend the linear models of the $R^2$ method by ranking each feature according to its statistical significance in a linear regression model. Development of the MIC was motivated by the difficulty of quantifying complex and nonlinear relationships [51]. The general premise is that the axes of a two-variable scatterplot can be discretized in such a way that each segment of an axis would share as many points as possible with a corresponding segment of the other; the optimum way to do this for any two variables defines the MIC value as a general measure of association. While there are some questions as to the perfect equitability of the MIC [52] – i.e., its agnosticism towards underlying function type – it has been implemented successfully for a number of practical applications where linear correlation may be insufficient [53]. Similarly, the Relief algorithm aims to produce a general measure of correlation by iteratively adjusting feature weights according to which features are best able to explain the differences or similarities between neighboring observations [54].

The filter-type feature selection methods were implemented in MATLAB R2019a [55], with the MIC values being computed through the external "minepy" library [56]. The wrapper-type methods were implemented through the "sequentialfs" function of MATLAB, using a linear support vector machine (SVM) to avoid overfitting the relatively small dataset. The wrapper method was run in both forward and backward modes: both adding features to a null list and removing them from a full list such that each step improves the model's cross validation score (5-fold selected herein) by the highest amount. To be more comparable to the final rankings output by the filter-type methods, a ranking of the wrapper results was constructed based purely on the relative frequency with which each feature shows up in the final list if the method is called 1000 times. Due to the random nature of how the function partitions the data for the model training, some fluctuation in the final list is expected, but increasing the number of iterations to 1500 did not significantly change the resulting frequency-based ranking.

## 3 Results and Discussion

### 3.1 First-principles calculations results

In general, with increasing imposed strain, the shear stress component first increases then, reaching a point of elastic instability, sharply decreases. The point of maximum stress before the drop is taken as the ideal shear strength of the alloy. A representative example of this pattern of shear stress evolution is shown in Figure 2 for the case of Ni-V (i.e., $Ni_{11}V$ in the present work). However, not all alloys studied show such a sudden drop in stress. Alloys containing Hf, Y, and Zr exhibit a more gradual slope from the maximum stress to the negative stress regime that follows. An example of this slower drop-off in stress is also shown in Figure 2 for the case of Ni-Zr. A slow drop-off in stress is notably present only in systems with a lower ideal shear strength, as is the case for the alloys with Hf, Y, and Zr. A decreased bond strength caused by the addition of these three elements could be responsible for both the low ideal shear strength and the more gradual shape of the stress profile.

The calculated ideal shear strengths are shown in Table 2. Previous calculations, also based on pure alias shear and using the same generalized gradient approximation approach [57], found the ideal shear strength of pure Ni to be 5.0 GPa [37], in good agreement with the 5.13 GPa found here. Experimental estimates of the idea shear strength of pure Ni are around 8 GPa [58], where the difference could be explained by the complex stress state imposed during nanoindentation [5]. Most ideal shear strengths of the alloys considered here fall within the range of 3.9 to 5.5 GPa, with the exceptions of Sc, Hf, Zr, and Y, which have lower values (< 3.7 GPa). The ideal shear strengths for all alloys are plotted against the atomic volume of the alloying element in Figure 3. Larger atomic volumes of the alloying elements are visibly associated with a lower ideal shear strength, in agreement with the calculated $R^2$ value of 0.839. Alloys with the four lowest ideal shear strengths also have the four largest atomic volumes. Some elements, like Si, Cu, and Ni itself, fall below the general trend of volume versus ideal shear strength. The explanation for this lies in a combination of features secondary in importance to volume, such as the Debye temperature, heat capacity, or electronic density used in the Miedema model [59], all of which show low values for Si, Cu, and Ni relative to other elements of similar size. Broad trends are difficult to spot by eye, however, and more rigorous correlational analysis is needed to reveal associations obscured by high dimensionality.

### 3.2 Correlational analysis results based on elemental properties

Several feature selection techniques were employed to better infer the physics underpinning the relationship between ideal shear strength and alloying element properties. The results of both filter and wrapper methods are listed in Table 3. There is a general agreement between the methods that among the most important features are atomic volume, radius, electronic density, electronegativity (both Pauling and Miedema variants), and some metric of position within the periodic table (group, period, or Mendeleev number). The strong association of Debye temperature to ideal shear strength, especially according to linear regression and the maximal information coefficient, is perhaps due to the information it contains about lattice site interactions and atomic bonding, but such specific conjectures cannot be directly supported by this general correlational analysis. Other features considered here, namely elastic and atomic properties, can be combined to estimate the Debye temperature [60]. However, doing so using these data shows a high variation in the predicted temperatures, which will not make for a performant feature. Therefore, within the present data, the experimental Debye temperature cannot be supplanted by a combination of other features.

Various electronic structure features are highly associated with ideal shear strength, which is in accordance with previous studies commenting on the importance of electron density and bond directionality during the shearing process [12]. More spherical valence charge density has been shown to ease the shearing process [61]. Other features, despite their ostensible similarities to the top-ranked features, are consistently found in low regard, such as the electron affinity.

It should be noted that the results of the wrapper methods shown in Table 3 discard much of the information obtained by following the sequential algorithm. Each iteration of the algorithm yields an optimized set of features that minimized the validation error of the model. If, for instance, the same critical information is encoded into two separate features, each optimal list might only contain one feature or the other. By only considering the overall frequency with which features appear after many iterations of the wrapper method, these two features will both be ranked at a 50% frequency, despite the information they contain being included in the model 100% of the time. As a result, features containing overlapping information that is important to the shearing process may have a lower frequency score than a moderately associated feature containing information found nowhere else in the feature set. As an example of this in the present data, the Young's and shear moduli were highly-ranked in the forward wrapper method results, with the bulk modulus lagging slightly behind, but usually only one or two of these, rarely all three, appeared in the final feature list of any individual iteration.

### 3.3 Correlational analysis results based on shear modulus of alloys

Although the shear modulus of the overall alloy is historically associated with ideal shear strength, as in Eqn. 1, not every correlation metric used here supports this for the shear modulus of the alloying elements themselves. Features that consistently rank highly include atomic volume, radius, and electronegativity, which have more in common with the framework of the Hume-Rothery rules of substitutional solid solutions (where the solute and solvent elements have a similar atomic size, crystal structure, valence, and electronegativity) than they do with the idea that shear modulus determines ideal shear strength [49]. However, there is a key difference between the shear modulus used in Eqn. 1 and the features discussed so far: all features listed in Table 1 are atomic features of the alloying elements in their pure bulk form, not properties of the overall alloys.

If instead the overall modulus of the alloy is considered for each Ni-X case, the shear modulus does indeed correlate well with ideal shear strength. Based on the applied deformation, the appropriate alloy shear modulus to compare against ideal shear strength is the $C_{55}$ component of the stiffness matrix. The relationship of ideal shear strength versus shear modulus is plotted in Figure 4 and shows a strongly linear behavior with a slope of 0.119 and $R^2$ = 0.975. Analytical forms of the ideal shear stress, using the fcc geometric factor $b/h = 1/\sqrt{2}$, predict a slope of $1/(\pi\sqrt{8}) = 0.1125$, in very close agreement with the slope found in the present work [8]. Further details of the linear fit for the case of both $C_{55}$ and the Hill average shear modulus, $G_{Hill}$, are listed in Table 4. Notably, the average shear modulus shows a lower slope, at 0.0886, which supports the use of a shear modulus specific to the deformation mode of interest, something not often considered when discussing ideal shear strengths in the general context of Eqn. 1.

Adding the shear modulus, $C_{55}$, of the alloys into the feature list and repeating the ranking procedure shows that $C_{55}$ of the alloy performs much better than other features. The rankings and association scores for this added feature are given in Table 5. The alloy shear modulus overwhelmingly appears as the best predictor of ideal shear strength, but it is fundamentally different than the rest of the features: atomic features are useful for arbitrary alloy compositions while finding the alloy shear modulus requires a separate calculation for each alloy of interest. In other words, when properties of each alloy are known, the alloy shear modulus is a great predictor of ideal shear strength, but when considering only existing knowledge of the separate alloying elements, the relative atomic size and electronic properties are most important.

## 4 Applications

### 4.1 Feature selection for mechanical property predictions

Recent insights into the application of machine learning on the small datasets characteristic of materials science problems highlight the benefit of effective features that improve model precision without increasing the number of degrees of freedom [62]. There is a need to identify key atomic features for mechanical property predictions to limit the dimensionality of future mechanical property predictions and improve model performance without drastically increasing the required input data quantity.

The results presented in Table 3 are specific to the Ni-X binary systems and may not perform well when implemented in machine learning applications on other alloy systems. However, the ideal shear strength data for dilute Ni-based alloys can be combined with similar datasets for other alloy systems. As an example, the results of the present study were combined with those of a first-principles study of 14 dilute Mg-based alloys of $Mg_{95}X$ compositions that investigated stacking and twinning fault energies in addition to ideal shear strengths [63]. The elements considered in the Mg-X alloys were Al, Ca, Cu, La, Li, Mn, Sc, Sn, Sr, Ti, Y, and Zr. For each alloying element, all atomic features listed in Table 1 were normalized by their value for the host element of Mg and added to the Ni-X dataset. The results of repeating the feature ranking procedure are given in Table 6.

The differences in feature importance between the Ni-X data and a more diverse dataset, that of the combined Ni- and Mg-based alloys, can be seen through a comparison of Table 3 and Table 6. In the combined data, elastic properties show a relatively greater importance; e.g., Young's modulus and shear modulus move up 12 and 14 positions, respectively, according to the r-Relief ranking and 6 and 3 positions according to $R^2$ values. In comparison to the Ni-alloy data, all association metrics except for the MIC show a decreased importance of the atomic volume, favoring instead features describing the electronic structure of the alloying element, such as electronic density or negativity values used in the Miedema model [59]. The combined Ni and Mg alloy results show that features describing position on the periodic table, especially Mendeleev number and group, remain consistently important to ideal shear strength.

In addition to changes in importance of atomic features, expanding the size of the dataset also changes the maximum association scores for each correlation method. For both the F-tests and MIC, the association score is greater with the combined data due to having a larger number of observations for each association. In contrast, the maximum association metrics for $R^2$ and the regression Relief algorithm decrease in the combined dataset due to the wider variety of data the methods are attempting to describe. The metrics shown for both wrapper methods are relative by definition, although it is notable that the forward wrapper method in the case of the combined data shows no single feature that is included in every iteration. On the Ni data, the forward wrapper gave scores of 1.000 and 0.488 to its first and second ranking features; with the combined data, the distribution is tight and shifted down in value, with the top two scores being 0.882 and 0.707, indicating the use of several features to predict what was before well-described by a single feature in the context of Ni-based alloys. Additionally, the forward wrapper method now ranks its previous top-choice, atomic volume, as $20^{th}$ in importance.

The shift in feature importance as the dataset expanded does not invalidate the findings on smaller datasets. As data from a wider variety of material systems is added into the feature selection procedure, it is expected that trends shift and become weaker as they become more general. For example, note that the association strength between the alloy ideal shear strength and the former top-ranked feature for the Ni alloy dataset, the atomic volume, has decreased for all association metrics. This could indicate either that the Mg alloy dataset has less of a dependence on the relative atomic volume of the alloying element or that the trends in atomic volume are not consistent across material systems due to differences in crystallography or other material properties. Those features shown to correlate well with Ni-X ideal shear strength may be more specific to the alloying environment of Ni than the features shown to be good predictors of ideal shear strength in both the Ni and Mg alloys. If the only goal is to create a model of Ni mechanical properties, only the Ni-X dataset should be used for training. However, as interest expands to creating a more general set of features for general mechanical property predictions, the ideal feature set should similarly expand to capture the more general trends across alloy systems. Potential future work using a more automated workflow for the calculation of ideal shear strengths, as outlined in Ref. [64] or as can be implemented in a general first-principles framework for automated computations (see Ref. [65] and www.dfttk.org), can allow for a wider range of data to be considered in the construction of a more general feature set for mechanical properties.

4.2 Prediction of Ni-X stress-strain response

The ideal shear strength has recently been shown to be a useful part of a multiscale approach to predict the macroscopic stress-strain response of single crystals through the crystal plasticity finite element method, as described in Ref. [7]. Following the same procedure by extending the ideal shear strengths listed in Table 2, the stress-influenced ideal shear strengths and elastic constants were calculated for dilute Ni-X alloys with common alloying elements: Fe, Co, Cr, V, Al, Ti, and Nb. Using the elastic constants of each alloy, the ideal strengths are converted through a modified Peierls-Nabarro equation to realistic stresses required for dislocation migration [19]. To be applicable to large strain regimes, the edge and screw components of the Peierls-Nabarro stress can be combined through a linear relationship of shear strain [7]:

$$\tau_c^{\alpha,es} = \left(1 - w\gamma^{\beta}\right)\tau_f^{edge} + w\gamma^{\beta}\tau_f^{screw} \tag{3}$$

where $\tau_c^{\alpha,es}$ is the combined flow stress as a function of $\gamma^{\beta}$, shear strain on slip system $\beta$, and the weighting coefficient $w$. The weighting coefficient is here assumed to be the same as was found for the case of pure Ni, 0.33. The hardening model used here is that of Peirce et al. [66], which proposes that the hardening rate of each slip system evolves according to the following relation:

$$h_{\alpha\beta} = \left[q + (1-q)\delta_{\alpha\beta}\right] \cdot \left[h_0 \operatorname{sech}^2 \left|\frac{h_0 \gamma}{\tau_s - \tau_0}\right|\right] \tag{4}$$

where $h_{\alpha\beta}$ is the hardening of slip system $\alpha$ as a result of slip on all systems $\beta$, $h_0$ is the initial hardening rate, $\tau_0$ is the initial slip system strength, $\tau_s$ is the slip system saturation stress, and $\gamma$ is the cumulative slip summed over all slip systems. Here, the latent hardening ratio was $q = 1.4$, and $\delta_{\alpha\beta}$ is the Kronecker delta, indicating that the entire first bracketed term is equal to 1 for self-hardening and 1.4 for latent hardening. For the present calculations, $\tau_s$, is assumed to retain its pure Ni value of 300 MPa, while $\tau_0$ and $h_0$ are obtained from the calculated Peierls-Nabarro stresses from Eqn. 3. Figure 5 shows a representative example of how the hardening model parameters are found based on first-principles data for the case of $Ni_{11}Fe$.

The hardening parameters for pure Ni determined in the present work ($\tau_0 = 8.8$ , $h_0 = 104$ MPa) agree well with those in the original implementation of this procedure for pure Ni ($\tau_0 = 9.4$ MPa, $h_0 = 120$ MPa) [7]. Differences in the pure Ni parameter sets may be explained by the slightly larger supercell, used here to accommodate the alloying elements. The difference may also be explained by the reduced number of pre-strains considered, as three instead of four values of pre-strain were considered in the present calculations of pure Ni and all Ni-X alloys. The slight differences between the two parameter sets for pure Ni show that this simplification is justified in light of the increased computational cost for each additional pre-strain value.

The geometric model for CPFEM was a chain of 100 hexahedral, full integration elements (type C3D8 in Abaqus [67]) stacked face-to-face along the loading direction. This model produces almost identical results as a 20,000-element realistic geometry of single crystal wire tension simulations, used in Ref. [7] for comparison to the experiments of Ref. [68]. Specifically, the 100-element model calculated an overall stress to within 0.5 MPa (0.25%) of the 20,000-element model at an engineering strain of 50% when using identical parameters but took only minutes to run instead of days. Since there is no inherent length scale in CPFEM, the geometries are indeed expected to give equivalent results since they both allow for lattice rotation during single crystal deformation. Using the 100-element geometric model and the hardening parameters predicted through first-principles methods, the large strain deformation response was predicted for uniaxial tension loading along the $[\bar{1}\ 5\ 10]$, $[\bar{1}\ 2\ 8]$, and [0 1 1] directions. These engineering stress-strain curves are shown in Figure 6 as a summary of the present DFT predictions of $\tau_0$ and $h_0$ for each CPFEM result, previous predictions of pure Ni stress-strain response [7], and experimental data of pure Ni tensile deformation [68,69].

As expected, predictions for all orientations show the same ordering of the relative strength of each alloy, but they differ in their agreement with available experimental data for pure Ni. For the $[\bar{1}\ 5\ 10]$ orientation, the disagreement is in the small strain regime where the crystal plasticity hardening model is unable to capture the more complicated staged hardening behavior at smaller strains without losing accuracy at larger strains, indicating the potential for future improvement of the model. Only the cases of Co and Fe alloying elements result in an increase in hardening behavior in comparison to pure Ni, with both alloys showing a higher initial stress, $\tau_0$, and initial hardening rate, $h_0$. Alloys with other elements result in stress-strain curves very similar to that of pure Ni, such as the case of Cr, or show a softening effect, such as Al, V, Ti, and Nb. While Co is known to lower the stacking fault energies of Ni alloys, so is Cr, and to almost the same degree [70]; therefore, the relative stacking fault energy does not completely explain the strength trends seen in Figure 6.

A continuum view of solute strengthening suggests that a solute with any size difference from the host element will generate a strengthening effect due to the interaction of the resulting strain field on dislocation motion [71]. Therefore, the softening effect seen in Figure 6 indicates that the present calculations do not support the traditional continuum elastic contribution to strengthening. Rather, the present calculations directly consider the electronic structure evolution during the shearing process, which is imperceptible from the view of continuum elasticity. Even in advanced empirical models of solute strengthening, differences in bonding characteristics are considered mainly through isotropic elastic properties [72]. Also note that the stress-strain responses are essentially ordered by atomic volume, suggesting that relative volume difference is important not only in the continuum view but also in the electronic structure evolution during the shearing process.

The overall mechanism of solid solution strengthening has previously been split into several categories: short-range effects of elastic property differences, long-range interactions of dislocation strain fields, and dislocation core structure effects [73]. Since no dislocations are explicitly considered, the results of the present work do not easily fit into any of these categories. Indeed, the perspective of ideal shearing is a nontraditional view on the longstanding subject of solute hardening. These results likely capture some of the short-range effects due to the consideration of the solute atom within the slip plane, and the pre-strains applied in the procedure used to find the hardening parameters are meant to emulate long-range strain fields, such as those encountered between dislocation segments during deformation. The degree of variation in the stress-strain curves of Figure 6 indicates that the present method captures a significant contribution to mechanical properties from the electronic structure on the slip plane during the shearing process – a contribution largely ignored in more traditional or phenomenological approaches.

## 5 Conclusions

The present study explored the variation in ideal shear stress that resulted from alloying in dilute Ni-X binary systems. The ideal shear strengths of 26 $Ni_{11}X$ alloys were calculated through first-principles based pure alias shear deformation, and quantitative correlational analysis explored the associations between strength values and elemental features. The utility of the ideal shear strength data was shown in its combination with other similar datasets for the design of machine learning feature sets for mechanical property applications as well as in its incorporation into multiscale deformation modeling. The main conclusions are as follows:

- The atomic properties of the alloying elements that showed the strongest associations with ideal shear strength of dilute Ni alloys described the alloying element's relative size (volume and radius), electronic properties (electronegativity and measures of electron density), and position on the periodic table (group and Mendeleev number). The elastic properties of the alloying element in their bulk states were often the next-strongest predictors of ideal shear strength.
- The shear modulus of binary alloys, which requires additional calculations for each alloy considered, shows a very high linear correlation with ideal shear strength, in accordance with the literature. The constant of proportionality of this linear relationship agrees with the traditional theory of ideal shear strength, with the average shear modulus showing more deviation than the deformation-specific shear modulus.
- Considering a combined ideal shear strength dataset of Mg and Ni alloys, the atomic features shifted relative to those considered most important in the Ni alloys. The most important features included electronegativity and electronic density values, with elastic properties rising in importance and atomic size falling.
- By adopting a recently-proposed predictive framework [7], the ideal shear strengths informed a crystal plasticity hardening model to give macroscale single crystal stress-strain responses for several Ni-X alloys, showing significant effects from the electronic structure considerations inherent to the method.

**Figures**

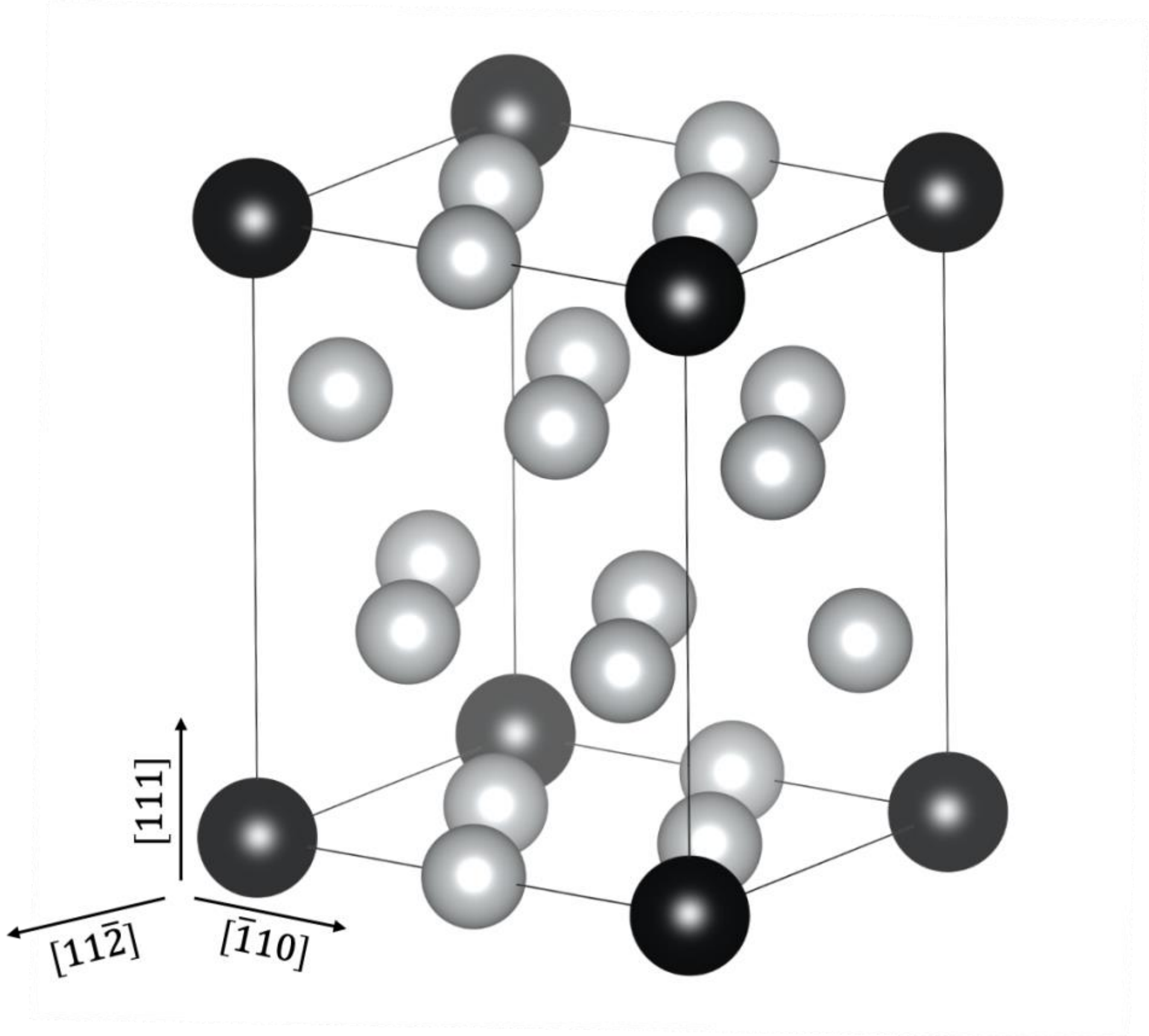

Figure 1. Orthorhombic supercell of rotated fcc lattice showing Ni atoms (gray) and those of the alloying element (black, corners), with lattice vectors shown. Although symmetry equivalent positions are shown in the figure, there are 12 unique atomic positions per supercell.

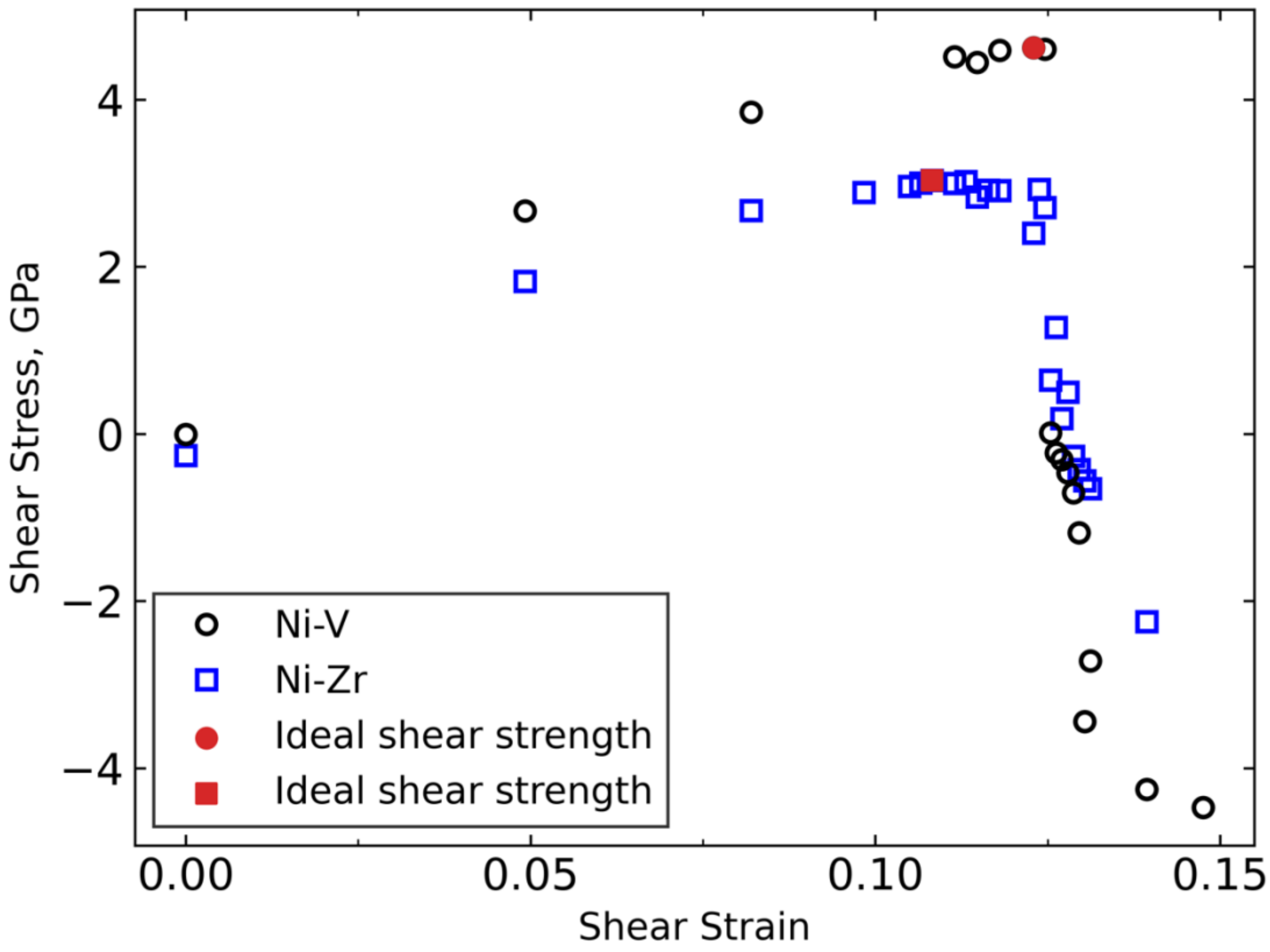

Figure 2. Calculated shear stress as a function of its imposed shear strain for the systems of Ni-V (black circles) and Ni-Zr (blue squares). Solid red symbols denote the ideal shear strength in both cases.

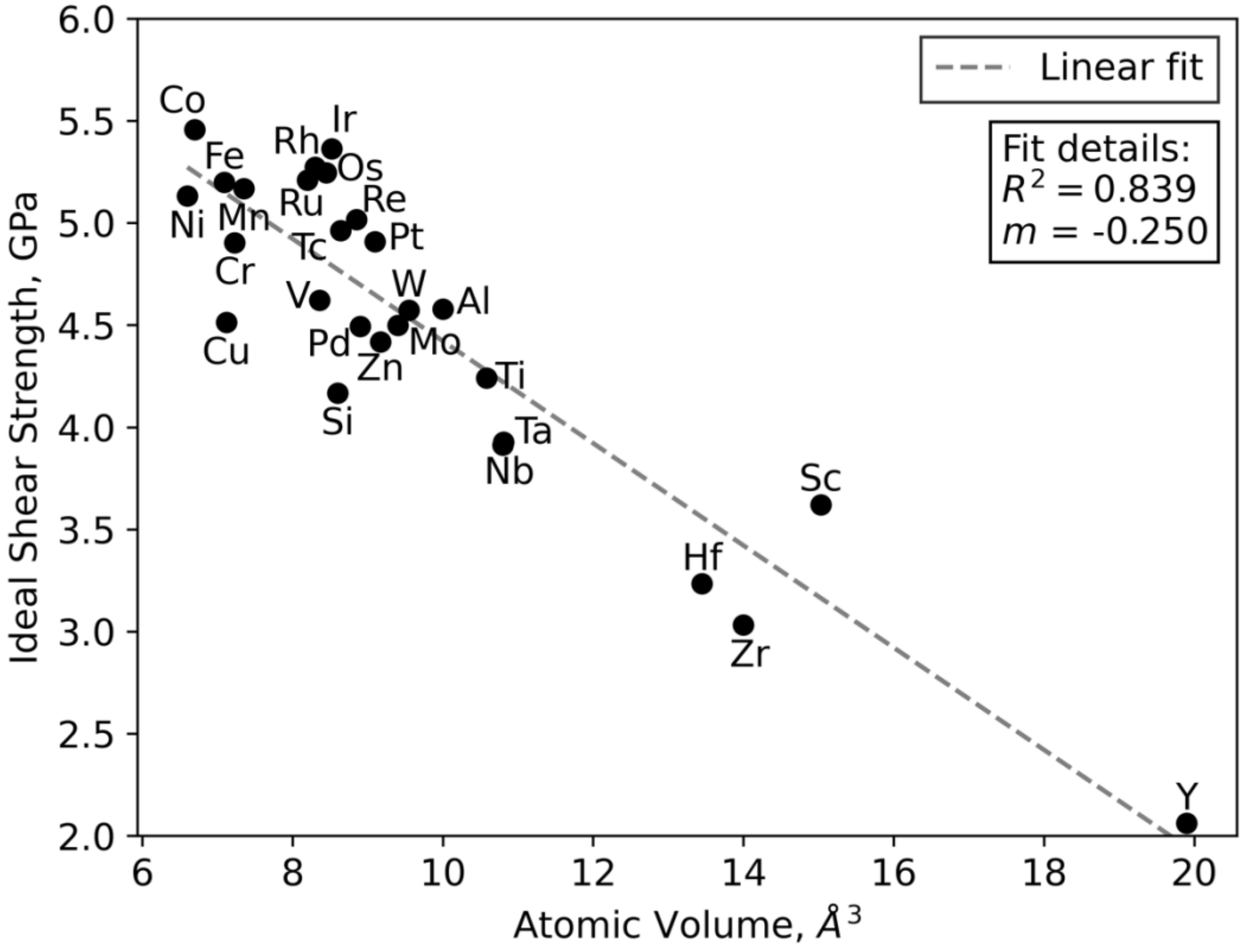

Figure 3. Ideal shear strength for 26 dilute Ni-based alloys and pure Ni plotted against the atomic volume with linear fit with coefficient of determination, $R^2$, and slope, $m$, included to highlight the linearity of the relationship.

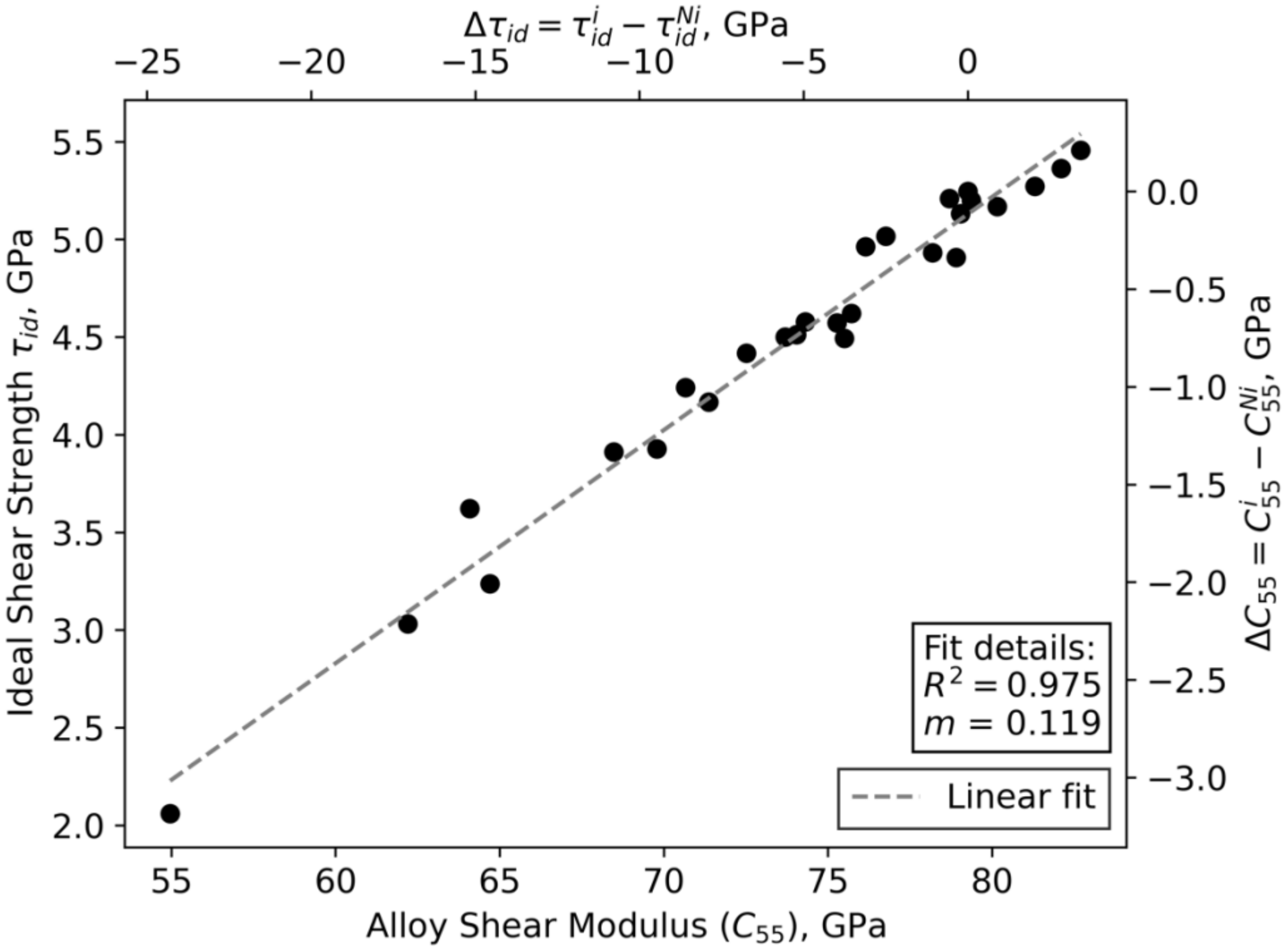

Figure 4. Relative ideal shear strength of the studied alloys plotted against relative $C_{55}$ value as a shear modulus, showing a highly linear relation, as seen in the details of the linear fit.

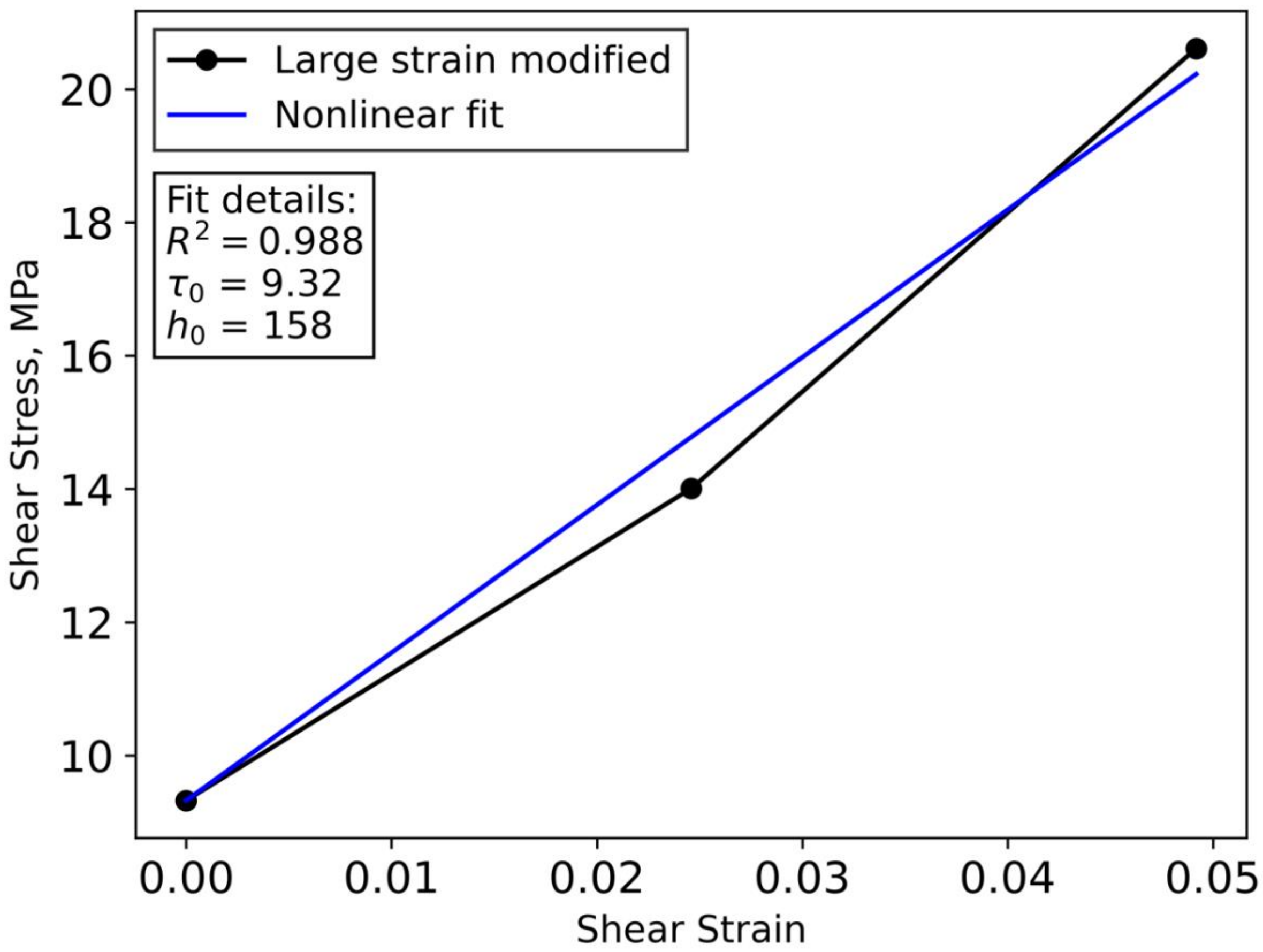

Figure 5. DFT-predicted stress-strain response of $Ni_{11}Fe$ modified for large strains through a model combining the effects of edge and screw dislocations based on the procedure described in Ref. [7], shown along with fitting parameters for the hardening model shown in Eqn. 4.

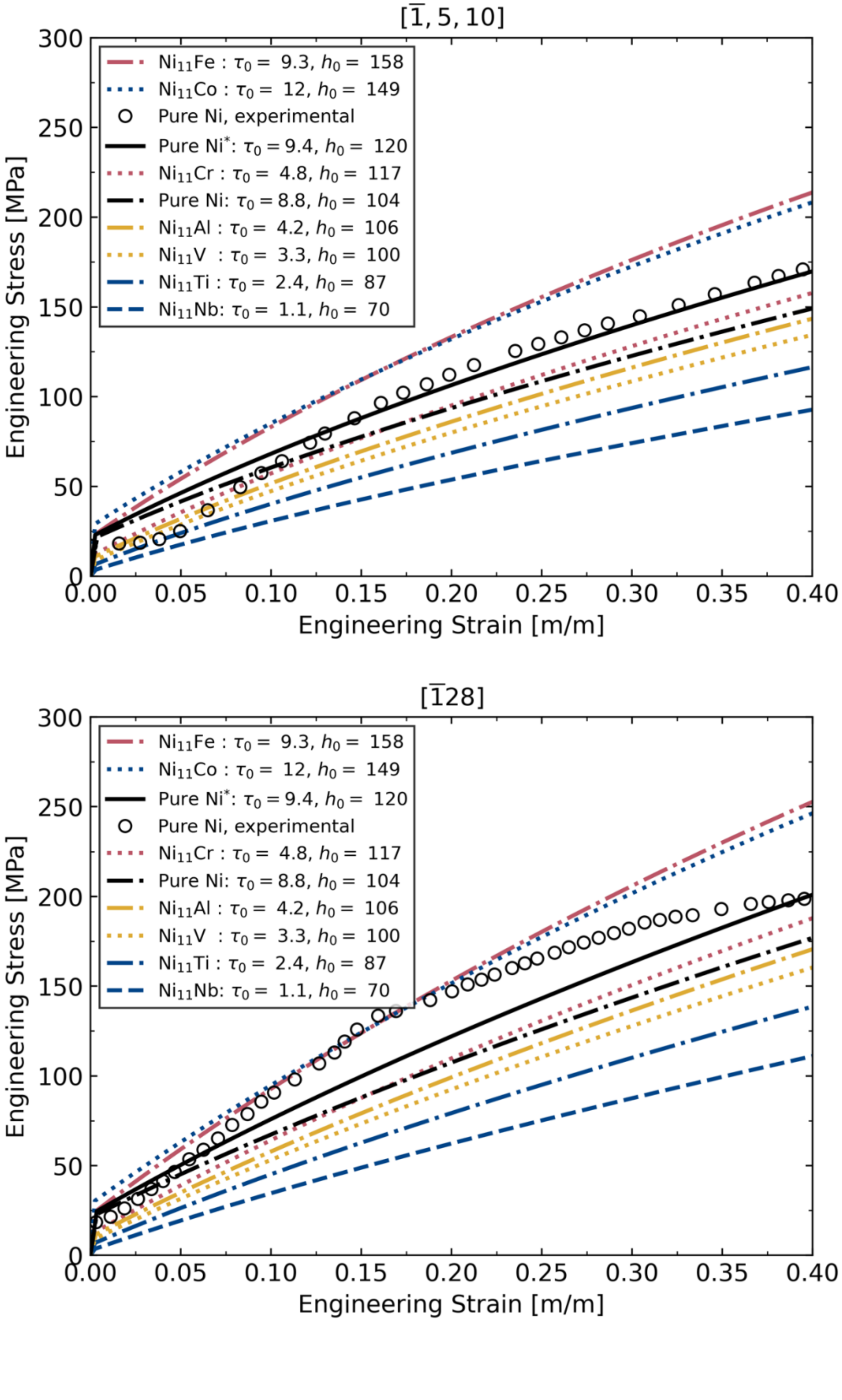
[$\bar{1}$, 5, 10]
Ni$_{11}$Fe : $\tau_0$ = 9.3, $h_0$ = 158
Ni$_{11}$Co : $\tau_0$ = 12, $h_0$ = 149
Pure Ni, experimental
Pure Ni*: $\tau_0$ = 9.4, $h_0$ = 120
Ni$_{11}$Cr : $\tau_0$ = 4.8, $h_0$ = 117
Pure Ni: $\tau_0$ = 8.8, $h_0$ = 104
Ni$_{11}$Al : $\tau_0$ = 4.2, $h_0$ = 106
Ni$_{11}$V : $\tau_0$ = 3.3, $h_0$ = 100
Ni$_{11}$Ti : $\tau_0$ = 2.4, $h_0$ = 87
Ni$_{11}$Nb: $\tau_0$ = 1.1, $h_0$ = 70
Engineering Stress [MPa]
Engineering Strain [m/m]
0
50
100
150
200
250
300
0.00
0.05
0.10
0.15
0.20
0.25
0.30
0.35
0.40
[$\bar{1}$28]
Ni$_{11}$Fe : $\tau_0$ = 9.3, $h_0$ = 158
Ni$_{11}$Co : $\tau_0$ = 12, $h_0$ = 149
Pure Ni*: $\tau_0$ = 9.4, $h_0$ = 120
Pure Ni, experimental
Ni$_{11}$Cr : $\tau_0$ = 4.8, $h_0$ = 117
Pure Ni: $\tau_0$ = 8.8, $h_0$ = 104
Ni$_{11}$Al : $\tau_0$ = 4.2, $h_0$ = 106
Ni$_{11}$V : $\tau_0$ = 3.3, $h_0$ = 100
Ni$_{11}$Ti : $\tau_0$ = 2.4, $h_0$ = 87
Ni$_{11}$Nb: $\tau_0$ = 1.1, $h_0$ = 70
Engineering Stress [MPa]
Engineering Strain [m/m]
0
50
100
150
200
250
300
0.00
0.05
0.10
0.15
0.20
0.25
0.30
0.35
0.40

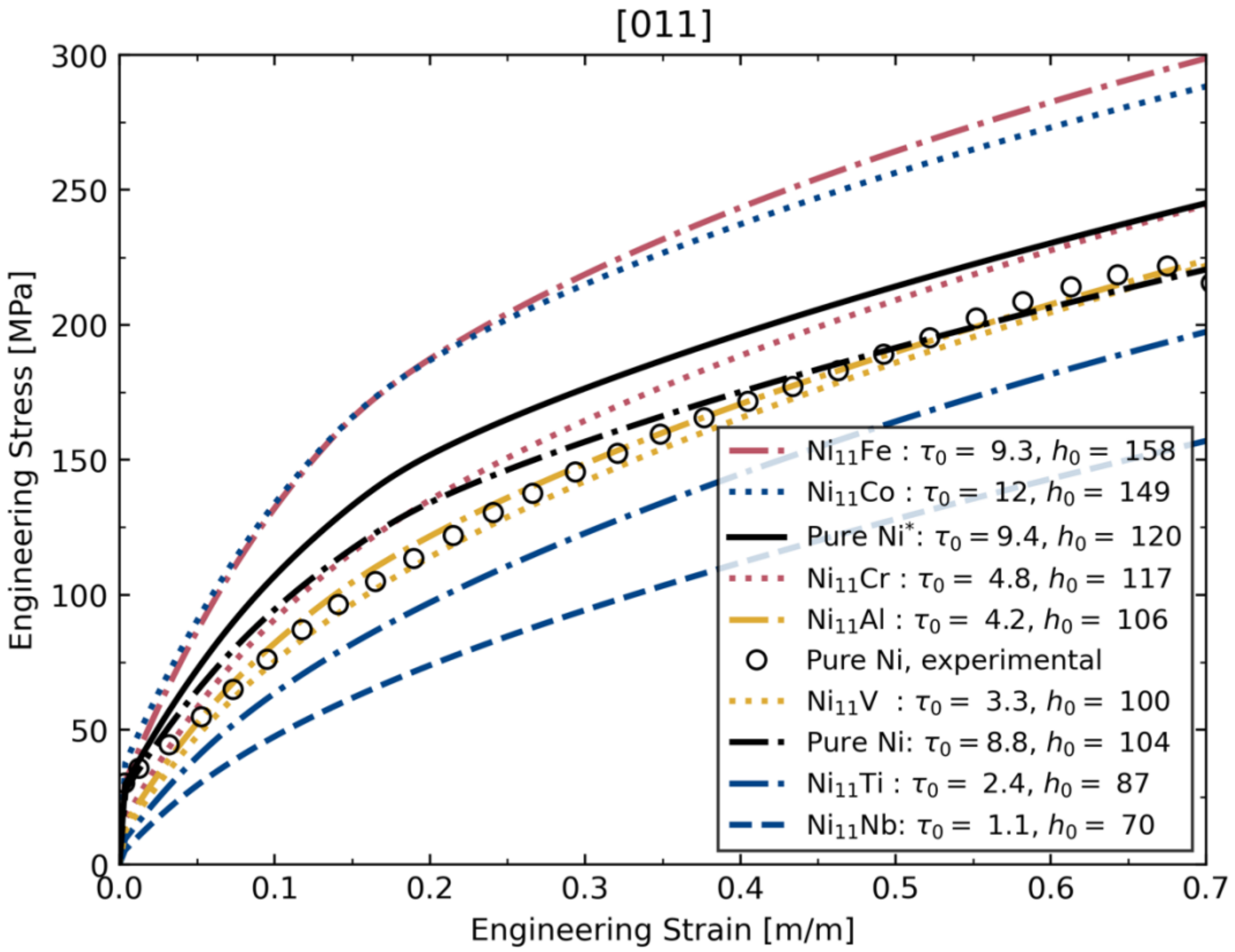

Figure 6. Engineering stress-strain curves for multiple dilute Ni alloy single crystals based on parameters (shown in legend) obtained through first-principles methods. Deformation was applied along the following orientations: $[\bar{1}\ 5\ 10]$ (top plot), $[\bar{1}\ 2\ 8]$ (middle), and $[0\ 1\ 1]$ (bottom). Experimental was data taken from Ref. [68] using a procedure detailed in Ref. [7]. The case of "Pure Ni*" indicates parameters from a previous study [7]. The legends are ordered by stress value at maximum strain.

## Tables

Table 1. List of features used along with brief explanations. Note that NfUnfill was not used here because all alloying elements had full or empty f-shells.

| | Feature | Explanation |
|---|---|---|
| Atomic | Radius_Coval | Covalent radius (pm) based on the collections of Wolfram Mathematica; see "ElementData" in Ref. [74]. |
| | Radius_vDW | Van der Waals atomic radius (pm) [74,75]. |
| | V0_Miedema | Atomic volume ($cm^3$/mol) used in the Miedema model [59]. |
| | Mass | Mass of pure elements. |
| Periodic | Group | Group of pure elements in the periodic table. |
| | M_Num2 | Mendeleev number MN2, starting bottom left and moving up then to the right [76][12]. |
| | Number | Atomic number of pure elements in the periodic table. |
| | Period | Period of pure elements in the periodic table. |
| Elastic | B | Bulk modulus (GPa) of pure elements based on [77,78]. Note that elastic properties of fcc Sr were taken from [79]. |
| | G | Shear modulus (GPa) of pure elements based on [77,78]. |
| | Y | Young's modulus (GPa) of pure elements based on [77,78]. |
| Thermodynamic | BoilingT | Boiling temperature (K) [80]. |
| | Ele_Conduc | Electrical conductivity of metals $(ohm\text{-}cm)^{-1}$ [81]. |
| | Heat_Capacity | Heat capacity at 298 K (J/kg-mol-K) [80,81]. |
| | Heat_Fusion | Heat of fusion at 298 K (J/mol) [81]. |
| | Heat_Sublimation | Heat of sublimation (J/mol) at 298 K [81]. |
| | MeltingT | Melting temperature (K) based on the collections by Kittel [81]. |
| | S298 | Standard entropy (J/mol-K) at 298 K [82]. |
| | Therm_Conduc | Thermal conductivity at 300 K (W/cm-K) [74,81]. |
| | VaporHeat | Vaporization heat (kJ/mol) based on the collections of Wolfram Mathematica; see "VaporizationHeat" in Ref. [74]. |
| Lattice | CohEnergy | Cohesive energy (eV/atom) collected by Kittel [81]. |
| | DebyeT | Debye temperature (K) collected by Kittel [81]. |
| | Va_Acti_FCC | Predicted vacancy activity energy of pure elements in the fcc structure, with the vacancy formation energy adopted for those with unstable fcc structures (i.e., Ge and La) [83]. |
| | Va_Form_FCC | Predicted vacancy formation energy of pure elements in fcc structure [83]. |
| Electronic | Electron_Affinity | Electron affinity (eV) [75]. |
| | EleDensity_Miedema | Electron density at the boundary of Wigner-Seitz cell used in the Miedema model [59]. |
| | EleNeg_Miedema | Electronegativity (volts) used in the Miedema model [59]. |
| | EleNeg_Pauling | Electronegativity (dimensionless) on the Pauling scale [74,75][13,16]. |
| | Ion_Pot_1 | The first ionization potential (eV) [80]. |
| | Ion_Pot_2 | The second ionization potential (eV) [80]. |
| | Ion_Pot_3 | The third ionization potential (eV) [80]. |
| | MaxR_Ele_in_Solid | Maximum range (mm) of electrons in solid elements for electron energy of 15 keV [80]. |
| | No_Spectral_lines | Number of spectral lines of the elements [80]. |
| | PPot_radius_s | Nonlocal pseudopotential radius (a.u.) for the s orbital [84]. |
| | PPot_radius_p | Nonlocal pseudopotential radius (a.u.) for the p orbital [84]. |
| | NsVal | Number of filled s-shell valence electron states. |
| | NpVal | Number of filled p-shell valence electron states. |
| | NdVal | Number of filled d-shell valence electron states. |
| | NfVal | Number of filled f-shell valence electron states. |
| | Nval | Number of filled valence electron states. |
| | NsUnfill | Number of unfilled s-shell valence electron states. |
| | NpUnfill | Number of unfilled p-shell valence electron states. |
| | NdUnfill | Number of unfilled d-shell valence electron states. |
| | NfUnfill | Number of unfilled f-shell valence electron states, not applicable here. |
| | Nunfill | Number of unfilled valence electron states. |

Table 2. Calculated ideal shear strengths for 26 alloying elements and pure Ni in units of GPa. The “pv” and “sv” listed after an atomic symbol indicates that the DFT calculations were performed while treating p or s states as valence states. Warmer colors indicate a higher value of ideal shear strength (blue=low, yellow=middle, red=high).

| | | | | | | | | | | | |
|---|---|---|---|---|---|---|---|---|---|---|---|
| | | | | | | | | | | Al | Si |
| | | | | | | | | | | 4.58 | 4.17 |
| Sc | Ti | V | Cr | Mn | Fe | Co | Ni, pure | Cu | Zn | | |
| 3.62 | 4.24 | 4.62 | 4.90 | 5.12 | 5.20 | 5.46 | 5.13 | 4.51 | 4.42 | | |
| Y_sv | Zr_sv | Nb_pv | Mo_pv | Tc_pv | Ru | Rh | Pd | | | | |
| 2.06 | 3.03 | 3.91 | 4.5 | 4.96 | 5.21 | 5.27 | 4.49 | | | | |
| | Hf_pv | Ta_pv | W_pv | Re | Os | Ir | Pt | | | | |
| | 3.23 | 3.93 | 4.57 | 5.02 | 5.24 | 5.36 | 4.91 | | | | |

Table 3. Feature ranking of Ni-X alloying elements based on their association with the calculated ideal shear strength. The association score of each method is shown in parentheses.

| Index | F-test | r-Relief | $R^2$ | MIC | Wrapper-forward | Wrapper-backward |
|---|---|---|---|---|---|---|
| 1 | Radius_Coval (16.354) | V0_Miedema ($1.22\times10^{-1}$) | V0_Miedema ($8.39\times10^{-1}$) | EleNeg_Miedema (0.709) | V0_Miedema (1.000) | V0_Miedema (1.000) |
| 2 | NdUnfill (11.907) | EleDensity_Miedema ($7.33\times10^{-2}$) | EleNeg_Miedema ($7.09\times10^{-1}$) | Y (0.669) | Y (0.488) | Heat_capacity (1.000) |
| 3 | Group (9.744) | M_Num2 ($6.93\times10^{-2}$) | EleDensity_Miedema ($5.88\times10^{-1}$) | G (0.669) | G (0.463) | Ele_Conduc (0.999) |
| 4 | EleNeg_Miedema (9.601) | Radius_Coval ($6.18\times10^{-2}$) | EleNeg_Pauling ($5.01\times10^{-1}$) | DebyeT (0.639) | Ele_Conduc (0.308) | EleNeg_Pauling (0.995) |
| 5 | NdVal (9.540) | EleNeg_Miedema ($4.98\times10^{-2}$) | Radius_Coval ($4.91\times10^{-1}$) | M_Num2 (0.621) | Heat_capacity (0.254) | Period (0.966) |
| 6 | V0_Miedema (9.379) | S298 ($4.69\times10^{-2}$) | DebyeT ($4.03\times10^{-1}$) | Radius_Coval (0.614) | MeltingT (0.241) | B (0.965) |
| 7 | NUnfill (8.901) | PPot_radius_s ($4.40\times10^{-2}$) | M_Num2 ($4.00\times10^{-1}$) | EleDensity_Miedema (0.593) | EleNeg_Miedema (0.237) | Radius_Coval (0.962) |
| 8 | EleNeg_Pauling (8.850) | PPot_radius_p ($3.51\times10^{-2}$) | NUnfill ($3.86\times10^{-1}$) | V0_Miedema (0.571) | Heat_Sublimation (0.218) | NpVal (0.959) |
| 9 | PPot_radius_p (8.826) | Max_range_electrons_in_solids ($3.11\times10^{-2}$) | G ($3.76\times10^{-1}$) | Group (0.542) | NpUnfill (0.157) | Number (0.953) |
| 10 | M_Num2 (8.481) | NdUnfill ($2.78\times10^{-2}$) | Y ($3.75\times10^{-1}$) | NUnfill (0.537) | Radius_Coval (0.132) | Nval (0.945) |
| 11 | Ion_Pot_3 (6.939) | Ion_Pot_3 ($2.47\times10^{-2}$) | B ($3.55\times10^{-1}$) | PPot_radius_p (0.536) | Va_Activa_FCC (0.116) | Y (0.943) |
| 12 | Y (6.795) | Period ($2.38\times10^{-2}$) | NdVal ($3.35\times10^{-1}$) | NdUnfill (0.535) | DebyeT (0.106) | NfVal (0.940) |
| 13 | PPot_radius_s (5.784) | DebyeT ($2.34\times10^{-2}$) | Ion_Pot_2 ($3.33\times10^{-1}$) | NdVal (0.532) | Atomic_radius_vDW | EleDensity_Miedema |
| | | | | | (0.095) | (0.938) |
| 14 | G (5.783) | Group $(2.06\times10^{-2})$ | NdUnfill $(3.06\times10^{-1})$ | Nval (0.484) | BoilingT (0.095) | Mass (0.932) |
| 15 | Number (4.966) | NdVal $(1.85\times10^{-2})$ | Ion_Pot_3 $(3.01\times10^{-1})$ | Heat_capacity (0.476) | EleDensity_Miedema (0.088) | NdUnfill (0.913) |
| 16 | Mass (4.702) | EleNeg_Pauling $(1.73\times10^{-2})$ | S298 $(2.59\times10^{-1})$ | EleNeg_Pauling (0.458) | VaporHeat (0.085) | Heat_Sublimation (0.903) |
| 17 | Ion_Pot_2 (4.489) | Ion_Pot_1 $(1.65\times10^{-2})$ | Group $(2.59\times10^{-1})$ | PPot_radius_s (0.458) | B (0.079) | EleNeg_Miedema (0.890) |
| 18 | EleDensity_Miedema (4.193) | Y $(1.42\times10^{-2})$ | Ion_Pot_1 $(2.02\times10^{-1})$ | B (0.454) | NsVal (0.077) | CohEnergy (0.888) |
| 19 | Nval (4.189) | NpVal $(1.35\times10^{-2})$ | PPot_radius_s $(1.45\times10^{-1})$ | Therm_Conduc (0.445) | EleNeg_Pauling (0.071) | DebyeT (0.887) |
| 20 | Vac_Form_FCC (3.723) | NUnfill $(1.15\times10^{-2})$ | Va_Activa_FCC $(1.41\times10^{-1})$ | Ele_Conduc (0.445) | Max_range_electrons_in_solids (0.058) | MeltingT (0.886) |
| 21 | Atomic_radius_vDW (3.631) | G $(6.89\times10^{-3})$ | Max_range_electrons_in_solids $(1.26\times10^{-1})$ | Ion_Pot_3 (0.414) | NpVal (0.058) | VaporHeat (0.873) |
| 22 | S298 (2.855) | NpUnfill $(6.19\times10^{-3})$ | No_Spectral_lines $(1.15\times10^{-1})$ | Electron_Affinity (0.390) | CohEnergy (0.054) | NdVal (0.861) |
| 23 | Electron_Affinity (2.847) | Atomic_radius_vDW $(5.84\times10^{-3})$ | PPot_radius_p $(1.08\times10^{-1})$ | Va_Activa_FCC (0.371) | NfVal (0.043) | Group (0.861) |
| 24 | DebyeT (2.749) | Heat_Sublimation $(5.17\times10^{-3})$ | Nval $(9.52\times10^{-2})$ | Ion_Pot_1 (0.361) | PPot_radius_p (0.040) | PPot_radius_s (0.860) |
| 25 | Therm_Conduc (2.650) | VaporHeat $(4.90\times10^{-3})$ | Therm_Conduc $(7.99\times10^{-2})$ | Ion_Pot_2 (0.361) | Electron_Affinity (0.035) | NUnfill (0.841) |
| 26 | Ele_Conduc (2.243) | Number $(1.95\times10^{-3})$ | Electron_Affinity $(7.77\times10^{-2})$ | No_Spectral_lines (0.316) | Heat_Fusion (0.031) | Heat_Fusion (0.835) |
| 27 | BoilingT (2.190) | Ion_Pot_2 $(1.38\times10^{-3})$ | Vac_Form_FCC $(3.68\times10^{-2})$ | S298 (0.308) | NsUnfill (0.026) | G (0.831) |
| 28 | Ion_Pot_1 (2.190) | B $(1.36\times10^{-3})$ | NsUnfill $(3.16\times10^{-2})$ | Max_range_electrons_in_solids (0.306) | Therm_Conduc (0.025) | PPot_radius_p (0.821) |
| 29 | Heat_capacity (2.031) | Heat_Fusion ($-1.45\times10^{-3}$) | Atomic_radius_vDW ($2.70\times10^{-2}$) | Number (0.297) | Ion_Pot_1 (0.023) | Ion_Pot_3 (0.812) |
| 30 | B (1.909) | CohEnergy ($-2.11\times10^{-3}$) | NsVal ($1.97\times10^{-2}$) | Mass (0.297) | Period (0.023) | Max_range_electrons_in_solids (0.779) |
| 31 | Max_range_electrons_in_solids (1.669) | Mass ($-2.37\times10^{-3}$) | Ele_Conduc ($9.14\times10^{-3}$) | BoilingT (0.279) | Group (0.023) | NpUnfill (0.758) |
| 32 | No_Spectral_lines (1.272) | Ele_Conduc ($-9.45\times10^{-3}$) | MeltingT ($6.47\times10^{-3}$) | Heat_Fusion (0.275) | Mass (0.020) | Ion_Pot_2 (0.714) |
| 33 | NsUnfill (0.980) | Va_Activa_FCC ($-1.12\times10^{-2}$) | NfVal ($5.74\times10^{-3}$) | Vac_Form_FCC (0.264) | PPot_radius_s (0.020) | Va_Activa_FCC (0.704) |
| 34 | Period (0.918) | Vac_Form_FCC ($-1.28\times10^{-2}$) | NpVal ($4.77\times10^{-3}$) | NsVal (0.244) | Number (0.019) | NsVal (0.696) |
| 35 | Va_Activa_FCC (0.676) | BoilingT ($-1.48\times10^{-2}$) | Heat_Fusion ($4.73\times10^{-3}$) | Atomic_radius_vDW (0.233) | M_Num2 (0.017) | BoilingT (0.679) |
| 36 | NpVal (0.394) | MeltingT ($-1.49\times10^{-2}$) | VaporHeat ($3.30\times10^{-3}$) | VaporHeat (0.213) | NUnfill (0.017) | S298 (0.678) |
| 37 | NsVal (0.388) | Heat_capacity ($-1.93\times10^{-2}$) | Mass ($3.22\times10^{-3}$) | NpVal (0.193) | NdUnfill (0.016) | NsUnfill (0.676) |
| 38 | NfVal (0.347) | No_Spectral_lines ($-1.94\times10^{-2}$) | Number ($3.06\times10^{-3}$) | NpUnfill (0.193) | Ion_Pot_2 (0.015) | M_Num2 (0.676) |
| 39 | Heat_Sublimation (0.291) | Nval ($-4.00\times10^{-2}$) | Heat_capacity ($2.65\times10^{-3}$) | CohEnergy (0.189) | NdVal (0.012) | Atomic_radius_vDW (0.670) |
| 40 | CohEnergy (0.164) | Therm_Conduc ($-4.40\times10^{-2}$) | Period ($2.25\times10^{-3}$) | Heat_Sublimation (0.189) | Nval (0.012) | Vac_Form_FCC (0.625) |
| 41 | Heat_Fusion (0.144) | Electron_Affinity ($-4.57\times10^{-2}$) | NpUnfill ($1.62\times10^{-3}$) | MeltingT (0.189) | Ion_Pot_3 (0.009) | Electron_Affinity (0.603) |
| 42 | VaporHeat (0.094) | NsVal ($-6.39\times10^{-2}$) | CohEnergy ($1.52\times10^{-3}$) | NsUnfill (0.187) | Vac_Form_FCC (0.007) | Ion_Pot_1 (0.530) |
| 43 | MeltingT (0.079) | NfVal ($-6.86\times10^{-2}$) | BoilingT ($9.20\times10^{-4}$) | Period (0.186) | S298 (0.006) | Therm_Conduc (0.499) |
| 44 | NpUnfill | NsUnfill | Heat_Sublimation | NfVal | No_Spectral_lines | No_Spectral_lines |

| | | | | | |
|---|---|---|---|---|---|
| (0.070) | ($-1.28\times10^{-1}$) | ($7.84\times10^{-4}$) | (0.111) | (0.003) | (0.393) |

Table 4. Details of the fitted linear dependence of the ideal shear strengths on deformation-specific shear modulus $C_{55}$ and averaged shear modulus $G_{Hill}$, with the proportionality constant from theory included for comparison.

| **Shear Modulus:** | **Slope:** | **Intercept:** | **$R^2$** |
|---|---|---|---|
| $C_{55}$ | 0.1194 | -4.333 | 0.975 |
| $G_{Hill}$ | 0.08728 | -2.89 | 0.985 |
| Theoretical | 0.1125 | 0 | 1 |

Table 5. Ranking of the shear modulus $C_{55}$ of each Ni-X alloy if the association methods had also considered this non-atomic feature along with the atomic features previously explored. Each association score is given in parentheses.

| **Method:** | **F-test** | **r-Relief** | **$R^2$** | **MIC** | **wrapper-forward** | **wrapper-backward** |
|---|---|---|---|---|---|---|
| **Score:** | 2nd (13.5) | 2nd (0.112) | 1st (0.975) | 1st (0.918) | 1st (1.00) | 1st (1.00) |

Table 6. Feature ranking of the combined Ni-X and Mg-X datasets based on their association with the calculated ideal shear strength. Mg-X ideal shear strength data was taken from ref [63], and the association score of each method is shown in parentheses.

| Index | F-test | r-Relief | $R^2$ | MIC | Wrapper-forward | Wrapper-backward |
|---|---|---|---|---|---|---|
| **1** | M_Num2 (21.754) | EleDensity_Miedema ($7.34\times10^{-2}$) | EleNeg_Miedema ($5.86\times10^{-1}$) | V0_Miedema (0.770) | EleNeg_Miedema (0.882) | Y (1.000) |
| **2** | EleNeg_Miedema (19.957) | EleNeg_Miedema ($5.81\times10^{-2}$) | EleDensity_Miedema ($5.75\times10^{-1}$) | G (0.714) | EleDensity_Miedema (0.707) | Heat_capacity (1.000) |
| **3** | Group (14.417) | M_Num2 ($4.45\times10^{-2}$) | EleNeg_Pauling ($4.19\times10^{-1}$) | M_Num2 (0.706) | Heat_capacity (0.702) | EleNeg_Miedema (0.999) |
| **4** | Therm_Conduc (13.774) | Group ($4.11\times10^{-2}$) | Y ($3.21\times10^{-1}$) | Y (0.700) | Y (0.458) | EleDensity_Miedema (0.999) |
| **5** | EleNeg_Pauling (12.750) | V0_Miedema ($3.33\times10^{-2}$) | B ($3.19\times10^{-1}$) | EleNeg_Miedema (0.639) | VaporHeat (0.458) | G (0.993) |
| **6** | G (11.679) | Y ($3.02\times10^{-2}$) | G ($3.17\times10^{-1}$) | Group (0.567) | MeltingT (0.408) | B (0.984) |
| **7** | Y (11.440) | G ($2.85\times10^{-2}$) | M_Num2 ($2.94\times10^{-1}$) | EleDensity_Miedema (0.553) | G (0.343) | CohEnergy (0.939) |
| **8** | Ele_Conduc (8.509) | Radius_Coval ($2.73\times10^{-2}$) | Radius_Coval ($2.74\times10^{-1}$) | EleNeg_Pauling (0.518) | Heat_Fusion (0.279) | Group (0.934) |
| **9** | EleDensity_Miedema (7.850) | EleNeg_Pauling ($2.57\times10^{-2}$) | Group ($2.60\times10^{-1}$) | Ion_Pot_3 (0.513) | Radius_Coval (0.274) | Heat_Fusion (0.929) |
| **10** | B (7.140) | PPot_radius_s ($2.45\times10^{-2}$) | Ion_Pot_2 ($2.47\times10^{-1}$) | No_Spectral_lines (0.504) | B (0.230) | MeltingT (0.929) |
| **11** | Ion_Pot_2 (6.523) | PPot_radius_p ($1.94\times10^{-2}$) | V0_Miedema ($2.34\times10^{-1}$) | Heat_capacity (0.490) | CohEnergy (0.225) | No_Spectral_lines (0.925) |
| **12** | PPot_radius_p (6.178) | B ($1.88\times10^{-2}$) | Max_range_electrons_in_solids ($2.11\times10^{-1}$) | Max_range_electrons_in_solids (0.462) | No_Spectral_lines (0.180) | VaporHeat (0.897) |
| **13** | V0_Miedema<br>(6.111) | Heat_capacity<br>($1.56\times10^{-2}$) | DebyeT<br>($1.79\times10^{-1}$) | Ele_Conduc<br>(0.438) | Group<br>(0.121) | M_Num2<br>(0.893) |
| **14** | PPot_radius_s<br>(5.948) | Ion_Pot_3<br>($1.37\times10^{-2}$) | Electron_Affinity<br>($1.75\times10^{-1}$) | Therm_Conduc<br>(0.414) | M_Num2<br>(0.116) | EleNeg_Pauling<br>(0.875) |
| **15** | Radius_Coval<br>(5.880) | Max_range_electrons_in_solids<br>($9.87\times10^{-3}$) | No_Spectral_lines<br>($1.35\times10^{-1}$) | PPot_radius_p<br>(0.406) | DebyeT<br>(0.111) | Number<br>(0.869) |
| **16** | No_Spectral_lines<br>(5.443) | S298<br>($4.42\times10^{-3}$) | Va_Activa_FCC<br>($1.10\times10^{-1}$) | Radius_Coval<br>(0.406) | BoilingT<br>(0.110) | S298<br>(0.869) |
| **17** | Ion_Pot_3<br>(5.348) | Heat_Sublimation<br>($4.25\times10^{-3}$) | Ion_Pot_3<br>($7.15\times10^{-2}$) | DebyeT<br>(0.402) | Ion_Pot_3<br>(0.098) | Mass<br>(0.864) |
| **18** | Va_Activa_FCC<br>(5.071) | Heat_Fusion<br>($2.15\times10^{-3}$) | S298<br>($6.46\times10^{-2}$) | PPot_radius_s<br>(0.396) | Heat_Sublimation<br>(0.096) | PPot_radius_p<br>(0.829) |
| **19** | Electron_Affinity<br>(4.525) | VaporHeat<br>($1.10\times10^{-3}$) | PPot_radius_s<br>($5.19\times10^{-2}$) | S298<br>(0.396) | Number<br>(0.049) | Radius_Coval<br>(0.817) |
| **20** | DebyeT<br>(3.905) | Period<br>($-3.39\times10^{-3}$) | Ion_Pot_1<br>($4.44\times10^{-2}$) | Ion_Pot_2<br>(0.392) | V0_Miedema<br>(0.048) | PPot_radius_s<br>(0.816) |
| **21** | Max_range_electrons_in_solids<br>(3.538) | Electron_Affinity<br>($-3.70\times10^{-3}$) | Vac_Form_FCC<br>($3.41\times10^{-2}$) | Electron_Affinity<br>(0.386) | EleNeg_Pauling<br>(0.047) | V0_Miedema<br>(0.810) |
| **22** | Vac_Form_FCC<br>(3.235) | CohEnergy<br>($-5.95\times10^{-3}$) | PPot_radius_p<br>(3.41×10-2) | B<br>(0.384) | Mass<br>(0.046) | Period<br>(0.806) |
| **23** | Heat_capacity<br>(2.955) | MeltingT<br>($-6.47\times10^{-3}$) | Atomic_radius_vDW<br>($3.16\times10^{-2}$) | Mass<br>(0.376) | Ele_Conduc<br>(0.043) | Va_Activa_FCC<br>(0.781) |
| **24** | Number<br>(2.836) | Va_Activa_FCC<br>($-8.87\times10^{-3}$) | BoilingT<br>($2.92\times10^{-2}$) | Va_Activa_FCC<br>(0.361) | S298<br>(0.035) | Max_range_electrons_in_solids<br>(0.746) |
| **25** | Heat_Fusion<br>(2.654) | No_Spectral_lines<br>($-9.74\times10^{-3}$) | Heat_capacity<br>($2.57\times10^{-2}$) | Ion_Pot_1<br>(0.331) | PPot_radius_p<br>(0.034) | BoilingT<br>(0.735) |
| **26** | Mass<br>(2.303) | BoilingT<br>($-1.07\times10^{-2}$) | CohEnergy<br>($1.62\times10^{-2}$) | Atomic_radius_vDW<br>(0.327) | PPot_radius_s<br>(0.025) | Atomic_radius_vDW<br>(0.674) |
| **27** | Heat_Sublimation (2.220) | Ion_Pot_2 ($-1.07\times10^{-2}$) | Heat_Sublimation ($8.37\times10^{-3}$) | Vac_Form_FCC (0.326) | Va_Activa_FCC (0.023) | Heat_Sublimation (0.629) |
| **28** | BoilingT (2.091) | DebyeT ($-1.45\times10^{-2}$) | Heat_Fusion ($5.58\times10^{-3}$) | Number (0.312) | Max_range_electrons_in_solids (0.019) | Electron_Affinity (0.628) |
| **29** | Ion_Pot_1 (1.699) | Mass ($-1.68\times10^{-2}$) | Therm_Conduc ($4.76\times10^{-3}$) | Heat_Fusion (0.305) | Period (0.011) | Ion_Pot_2 (0.625) |
| **30** | Atomic_radius_vDW (0.799) | Vac_Form_FCC ($-1.70\times10^{-2}$) | Number ($4.14\times10^{-3}$) | MeltingT (0.273) | Ion_Pot_1 (0.011) | Ele_Conduc (0.596) |
| **31** | S298 (0.489) | Number ($-1.80\times10^{-2}$) | Period ($2.50\times10^{-3}$) | BoilingT (0.264) | Electron_Affinity (0.010) | Ion_Pot_3 (0.537) |
| **32** | VaporHeat (0.306) | Ele_Conduc ($-2.19\times10^{-2}$) | MeltingT ($2.24\times10^{-3}$) | Heat_Sublimation (0.257) | Atomic_radius_vDW (0.006) | Therm_Conduc (0.477) |
| **33** | Period (0.133) | Ion_Pot_1 (-2.72×10-2) | VaporHeat ($1.82\times10^{-3}$) | Period (0.230) | Vac_Form_FCC (0.004) | Ion_Pot_1 (0.444) |
| **34** | MeltingT (0.083) | Atomic_radius_vDW ($-2.76\times10^{-2}$) | Mass ($1.57\times10^{-3}$) | VaporHeat (0.219) | Ion_Pot_2 (0.002) | Vac_Form_FCC (0.406) |
| **35** | CohEnergy (0.012) | Therm_Conduc ($-3.13\times10^{-2}$) | Ele_Conduc (7.94×10-4) | CohEnergy (0.189) | Therm_Conduc (0.001) | DebyeT (0.391) |

**Acknowledgements**

This work was financially supported by the U. S. Department of Energy (DOE) via award nos. DE-FE0031553 and DE-AR0001435. Computations were performed partially on The Pennsylvania State University's Institute for Computational and Data Sciences' Roar supercomputer, partially on the resources of NERSC supported by the DOE Office of Science under contract no. DE-AC02-05CH11231, and partially on the resources of XSEDE supported by NSF via grant no. ACI-1548562. JDS acknowledges support from the Department of Energy National Nuclear Security Administration Stewardship Science Graduate Fellowship, provided under cooperative agreement number DE-NA0003960.

**Data Availability**

The input files and raw data required to reproduce these findings are available to download from https://doi.org/10.5281/zenodo.5497912. The processed data are available to download from the supplemental information.